\documentclass[12pt]{JHEP3}

\title{On Entropy Function for Supersymmetric Black Rings}
\author{Rong-Gen Cai \\
        Institute of Theoretical Physics\\
    Chinese Academy of Sciences\\
    P.O.Box 2735, Beijing 100080, China\\
    \email{cairg@itp.ac.cn}}
\author{Da-Wei Pang\\
        Institute of Theoretical Physics\\
    Chinese Academy of Sciences\\
    P.O.Box 2735, Beijing 100080, China\\
    {\rm and}\\
    Graduate University of the Chinese Academy of Sciences\\
    YuQuan Road 19A, Beijing 100049, China\\
    \email{pangdw@itp.ac.cn}}

\abstract{The entropy function for five-dimensional supersymmetric
black rings, which are solutions of $U(1)^{3}$ minimal supergravity,
is calculated via both on-shell and off-shell formalism. We find
that at the tree level, the entropy function obtained from both
perspectives can reproduce the Bekenstein-Hawking entropy. We also
compute the higher order corrections to the entropy arising form
five-dimensional Gauss-Bonnet term as well as supersymmetric $R^{2}$
completion respectively and compare the results with previous
microscopic calculations.
 }

\keywords{Black Holes in String Theory; Black Holes}
\begin{document}

\section{Introduction}
\label{Introduction}
The attractor mechanism has played an important role in
understanding black hole physics in string theory and has been
studied extensively in the past decade. It was initiated in the
context of $N=2$ extremal black holes~\cite{classic} and
generalized to more general cases, such as supersymmetric black
holes with higher order corrections~\cite{cardoso} and
non-supersymmetric attractors~\cite{fgk},~\cite{non},~\cite{dst}.

Recently, based on Wald's entropy formula~\cite{wald}, A.Sen
proposed an effective method for calculating the entropy of
D-dimensional black holes with near horizon geometry $AdS_{2}\times
S^{D-2}$, which is named as ``entropy function'' method~\cite{sen}.
It states that the entropy of such kind of black holes can be
obtained by extremizing the ``entropy function'' with respect to
various moduli, where the entropy function is defined as integrating
the Lagrangian over the horizon coordinates and taking the Legendre
transformation with respect to the electric charges. This method has
been applied to many specific examples, such as extremal black holes
in higher dimensions, rotating black holes and non-supersymmetric
black holes. For recent developments, see~\cite{relate}. It is also
an useful way to calculate the higher order corrections to the
entropy. In particular, more recently we have shown that for some
nonextremal black holes in string theory, the entropy function
method also works quite well~\cite{CaiPang}. Similar arguments that
concerning entropy function for non-extremal black holes/branes
appeared very recently~\cite{gagh}.

It is well known that for black holes in four-dimensional
asymptotically flat spacetime, there exists only one horizon
topology $S^{2}$. But for black holes in five-dimensional
spacetime, the horizon topology is not unique. A black hole
solution with horizon topology $S^{1}\times S^{2}$, named as black
ring, was presented firstly in~\cite{er}. Several important
developments are listed
in~\cite{eemr1},~\cite{eemr2},~\cite{gg},~\cite{bena},~\cite{cgms},
where various solutions, the microscopic entropy and relations to
other topics are discussed. For reviews, see~\cite{rev}.

The near horizon geometry of certain black holes and black rings
turns out to be $AdS_{3}\times S^{2}$. It becomes $AdS_{2}\times
S^{2}$ after dimensional reduction so that one expects the entropy
function formalism also works. Such attempts have been discussed
in~\cite{diiss,cop} and an interesting paper appeared very
recently~\cite{gj}, where the entropy function for five-dimensional
extremal black holes and black rings is constructed in the context
of two-derivative gravity coupled to abelian gauge fields and
neutral scalar fields, which shows the validity of the entropy
function in a general way.

Since the entropy function method can give the higher order
corrections to the entropy conveniently, it is interesting to study
the corrections via the entropy function and compare the results
with microscopic calculations. In order to study the higher order
corrections to the entropy, we take supersymmetric black rings in
$U(1)^{3}$ supergravity as a concrete example. Firstly we carry out
the analysis using the two-derivative on-shell and off-shell
supergravity, where we make a dimensional reduction so that the
resulting action is gauge and coordinate invariant. We find that the
entropy function can reproduce both the Bekenstein-Hawking entropy
and the near horizon geometry, while the correct attractor values of
the moduli fields can also be obtained by extremizing the entropy
function. Then we calculate the $R^{2}$ corrections to black ring
entropy by adding the five-dimensional Gauss-Bonnet term which
originates from $R^{4}$ terms in M-theory compactified on a
Calabi-Yau manifold as well as supersymmetric $R^{2}$ completion and
compare our results with previous microscopic considerations.

The rest of the paper is organized as follows. In
section~\ref{SBR} we review the supersymmetric black ring
solutions in the $U(1)^{N}$ theory and specialize to the case of
$N=3$. After dimensional reduction to four-dimensional spacetime,
the entropy function for $U(1)^{3}$ black rings is carried out in
section~\ref{Four-dim}. The higher order corrections to the
entropy are discussed in section~\ref{HOC}.  We summarize the
results and discuss some related topics in section~\ref{Summary}.

\section{Supersymmetric Black Rings in $U(1)^{3}$ Theory}
\label{SBR}
In this section, we review some salient properties of
supersymmetric black ring in the $U(1)^{3}$ theory, which are
needed in the following calculations. For more details, see Sec.
II and Appendix B of~\cite{eemr2}.

Consider the case of minimal supergravity coupled to $N-1$ Abelian
vector multiplets with scalars taking values in a symmetric space.
The action for such a theory is
\begin{eqnarray}
\label{2eq1}
S&=&\frac{1}{16\pi
G_{5}}\int(R\star1-G_{IJ}dX^{I}\wedge\star
dX^{J}-G_{IJ}F^{I}\wedge\star F^{J}\nonumber\\
 & &-\frac{1}{6}C_{IJK}A^{I}\wedge F^{J}\wedge F^{K}),
\end{eqnarray}
where $I,J,K=1,\cdots,N$ and the constants $C_{IJK}$ are symmetric
in $(IJK)$. The $N-1$ dimensional scalar manifold is conveniently
parameterized by the $N$ scalars $X^{I}$, which obey the constraint
\begin{equation}
\frac{1}{6}C_{IJK}X^{I}X^{J}X^{K}=1.
\end{equation}
The matrix $G_{IJ}$ is defined by
\begin{equation}
G_{IJ}\equiv\frac{9}{2}X_{I}X_{J}-\frac{1}{2}C_{IJK}X^{K},
\end{equation}
where $X_{I}\equiv\frac{1}{6}C_{IJK}X^{J}X^{K}$ such that
$X_{I}X^{I}=1$.

The supersymmetric black ring in $U(1)^{3}$ theory can be viewed
as an eleven-dimensional supertube carrying three charges and
three dipoles after dimensional reduction down to $D=5$ on
$T^{6}$. The configuration can be summarized as follows:
\begin{eqnarray}
Q_{1}~~~M2:&1~2~-~-~-~-~-,\nonumber\\
Q_{2}~~~M2:&-~-~3~4~-~-~-,\nonumber\\
Q_{3}~~~M2:&-~-~-~-~5~6~-,\nonumber\\
p_{1}~~~m5:&-~-~3~4~5~6~\psi,\nonumber\\
p_{2}~~~m5:&1~2~-~-~5~6~\psi,\nonumber\\
p_{3}~~~m5:&1~2~3~4~-~-~\psi.
\end{eqnarray}
Such a configuration can be taken as a solution of $D=11$
supergravity with the effective action
\begin{eqnarray}
S_{11}&=&\frac{1}{16\pi
G_{11}}\int(R_{11}\star_{11}1-\frac{1}{2}\mathcal{F}\wedge\star_{11}\mathcal{F}\nonumber\\
      & &-\frac{1}{6}\mathcal{F}\wedge\mathcal{F}\wedge\mathcal{A}).
\end{eqnarray}
The eleven-dimensional solution describing this system takes the
form
\begin{eqnarray}
\label{2eq6}
ds_{11}^{2}&=&ds_{5}^{2}+X^{1}(dz_{1}^{2}+dz_{2}^{2})+X^{2}(dz_{3}^{2}+dz_{4}^{2})\nonumber\\
           & &+X^{3}(dz_{5}^{2}+dz_{6}^{2}),\nonumber\\
\mathcal{A}&=&A^{1}\wedge dz_{1}\wedge dz_{2}+A^{2}\wedge
dz_{3}\wedge dz_{4}\nonumber\\
           & &+A^{3}\wedge dz_{5}\wedge dz_{6},
\end{eqnarray}
where $z_{i}$ denote the coordinates along the 123456-directions
and $\mathcal{A}$ is the three-form potential.

Note that if we reduce the eleven-dimensional action to
five-dimensional spacetime on $T^{6}$ using the ansatz~(\ref{2eq6}),
we will obtain precisely the action~(\ref{2eq1}) with $N=3$,
$C_{IJK}=1$ if $(IJK)$ is a permutation of $(123)$ and $C_{IJK}=0$
otherwise, and
\begin{equation}
G_{IJ}=\frac{1}{2}{\rm diag}[(X^{1})^{-2}, (X^{2})^{-2},
(X^{3})^{-2}].
\end{equation}
The resulting five-dimensional black ring solution is
characterized by the metric $ds_{5}^{2}$, three scalars $X^{i}$,
and three one-forms $A^{i}$, with field strengths $F^{i}=dA^{i}$,
which are given by
\begin{eqnarray}
ds_{5}^{2}&=&-(H_{1}H_{2}H_{3})^{-2/3}(dt+\omega)^{2}+(H_{1}H_{2}H_{3})^{1/3}d{\bf
x}_{4}^{2},\nonumber\\
A^{i}&=&H_{i}^{-1}(dt+\omega)-\frac{p_{i}}{2}[(1+y)d\psi+(1+x)d\phi],\nonumber\\
X^{i}&=&H_{i}^{-1}(H_{1}H_{2}H_{3})^{1/3},
\end{eqnarray}
where
\begin{eqnarray}
d{\bf
x}_{4}^{2}&=&\frac{R^{2}}{(x-y)^{2}}[\frac{dy^{2}}{y^{2}-1}+(y^{2}-1)d\psi^{2}\nonumber\\
          &
          &+\frac{dx^{2}}{1-x^{2}}+(1-x^{2})d\phi^{2}],\\
H_{1}&=&1+\frac{Q_{1}-p_{2}p_{3}}{2R^{2}}(x-y)-\frac{p_{2}p_{3}}{4R^{2}}(x^{2}-y^{2}),\nonumber\\
H_{2}&=&1+\frac{Q_{2}-p_{1}p_{3}}{2R^{2}}(x-y)-\frac{p_{1}p_{3}}{4R^{2}}(x^{2}-y^{2}),\\
H_{3}&=&1+\frac{Q_{3}-p_{1}p_{2}}{2R^{2}}(x-y)-\frac{p_{1}p_{2}}{4R^{2}}(x^{2}-y^{2}),\nonumber\\\nonumber
\end{eqnarray}
and $\omega=\omega_{\phi}d\phi+\omega_{\psi}d\psi$ with
\begin{eqnarray}
\omega_{\phi}&=&-\frac{1}{8R^{2}}(1-x^{2})[p_{1}Q_{1}+p_{2}Q_{2}+p_{3}Q_{3}\nonumber\\
             & &-p_{1}p_{2}p_{3}(3+x+y)],\nonumber\\
\omega_{\psi}&=&\frac{1}{2}(p_{1}+p_{2}+p_{3})(1+y)-\frac{1}{8R^{2}}(y^{2}-1)\nonumber\\
             &
             &\times[p_{1}Q_{1}+p_{2}Q_{2}+p_{3}Q_{3}-p_{1}p_{2}p_{3}(3+x+y)].
\end{eqnarray}
Note that the six-torus $T^{6}$ has constant volume because
$X^{1}X^{2}X^{3}=1$.

The horizon locates at $y=-\infty$ and in order to obtain the near
horizon geometry, we have to take rather complicated coordinate
transformations, which are discussed extensively in Appendix D
of~\cite{eemr2} and here we will not repeat any more. The
resulting near horizon metric is
\begin{eqnarray}
\label{2eq12}
ds^{2}&=&\frac{4L}{p}\tilde{r}d\tilde{t}d\tilde{\psi}+L^{2}d\tilde{\psi}^{2}+
\frac{p^{2}}{4}\frac{d\tilde{r}^{2}}{\tilde{r}^{2}}\nonumber\\
      & &+\frac{p^{2}}{4}(d\theta^{2}+\sin^{2}\theta d\phi^{2}),
\end{eqnarray}
where
\begin{eqnarray}
L&\equiv&\frac{1}{2p^{2}}[2\sum\limits_{i<j}\mathcal{Q}_{i}p_{i}\mathcal{Q}_{j}p_{j}-
\sum\limits_{i}\mathcal{Q}_{i}^{2}p_{i}^{2}-4R^{2}p^{3}\sum\limits_{i}p_{i}],\nonumber\\
\mathcal{Q}_{1}&=&Q_{1}-p_{2}p_{3},~~~\mathcal{Q}_{2}=Q_{2}-p_{1}p_{3},~~~\mathcal{Q}_{3}=Q_{3}-p_{1}p_{2},\nonumber\\
p&\equiv&(p_{1}p_{2}p_{3})^{1/3}.
\end{eqnarray}
Finally, let $\tilde{t}=p^{2}\tau/4$ and $e^{0}=p/2L$, the near
horizon metric~(\ref{2eq12}) becomes
\begin{eqnarray}
ds^{2}&=&\frac{p^{2}}{4}(-\tilde{r}^{2}d\tau^{2}+\frac{d\tilde{r}^{2}}{\tilde{r}^{2}})+L^{2}(d\tilde{\psi}
   +e^{0}\tilde{r}d\tau)^{2}\nonumber\\
      & &+\frac{p^{2}}{4}(d\theta^{2}+\sin^{2}\theta d\phi^{2}),
\end{eqnarray}
which is the product of a locally $AdS_{3}$ with radius $p$ and a
two-sphere of radius $p/2$. The Bekenstein-Hawking entropy is
\begin{equation}
\label{2eq15}
S_{BH}=\frac{\mathcal{A}_{5}}{4G_{5}}=\frac{2\pi^2Lp^{2}}{4G_{5}}.
\end{equation}
\section{The Entropy Function Analysis in Four-dimensional Spacetime}
\label{Four-dim}
In this section, we will carry out the analysis of entropy function
for supersymmetric black rings in detail, using both the on-shell
and off-shell Lagrangian at two-derivative level.
\subsection{On-Shell Analysis}
\label{on}
In this subsection, we calculate the entropy of supersymmetric black
rings in $U(1)^{3}$ supergravity via the entropy function formalism,
which can be seen as a concrete example of~\cite{gj}. According
to~\cite{sen}, in order to carry out entropy function analysis, the
Lagrangian must be gauge and coordinate invariant. Since the
five-dimensional effective action contains a Chern-Simons term, the
Lagrangian is not gauge invariant and we have to reduce it to four
dimensions so that the entropy function can be applied. Such
analysis was initiated in~\cite{ss1} and has been followed
in~\cite{ss2},~\cite{cop} and~\cite{gj}.

We take the near horizon field configuration as follows:
\begin{eqnarray}
\label{3eq1} &
&ds_{5}^{2}=w^{-1}[v_{1}(-r^{2}dt^{2}+dr^{2}/r^{2})+v_{2}(d\theta^{2}
+\sin^{2}\theta d\phi^{2})]+w^{2}(d\psi+e^{0}rdt)^{2},\nonumber\\
& &A^{I}_{5}=A^{I}_{4}+a^{I}(d\psi+e^{0}rdt), \nonumber\\
&
&F^{I}_{5rt}=e^{I}+a^{I}e^{0},~~~F^{I}_{4rt}=e^{I},~~~F^{0}_{4rt}=e^{0},
~~~F^{I}_{5\theta\phi}=F^{I}_{4\theta\phi}=\frac{1}{2}p^{I}\sin\theta,\nonumber\\
& &X^{I}=x^{I},~~~I=1,2,3.
\end{eqnarray}
Note that the $\psi$ components of the gauge potential become axions
in four-dimensional spacetime. The entropy function analysis will be
processed in four-dimensional spacetime after dimensional reduction
on $\psi$ coordinate. Define
\begin{equation}
\label{3eq2} f_{0}\equiv\frac{1}{16\pi}\int d\theta
d\phi\sqrt{-g}(\mathcal{L}^{'}_{0}+\mathcal{L}_{0CS})=f^{'}_{0}+f_{0CS},
\end{equation}
where $\mathcal{L}_{0CS}$ is the Chern-Simons term down to four
dimensions and $\mathcal{L}^{'}_{0}$ denotes the resulting
Lagrangian coming from the gauge and coordinate invariant terms in
the original five-dimensional supergravity. Throughout the work, the
four-dimensional Newton constant is set to be $G_{4}\equiv1$ so that
$G_{5}=2\pi$.

Note that for a consistent dimensional reduction, the first term
in~(\ref{3eq2}) can be evaluated in the original five-dimensional
background without writing out the expression for the reduced action
explicitly, what we should pay attention to is the second
Chern-Simons term. In four-dimensional spacetime such a term becomes
\begin{eqnarray}
\frac{1}{6}A_{5}\wedge F_{5}\wedge
F_{5}=e^{-1}(\frac{1}{6}C_{IJK}a^{I}a^{J}a^{K}F^{0}_{4\mu\nu}F^{0}_{4\lambda\sigma}
\epsilon^{\mu\nu\lambda\sigma}+
\frac{1}{4}C_{IJK}a^{J}a^{K}F^{I}_{4\mu\nu}F^{0}_{4\lambda\sigma}
\epsilon^{\mu\nu\lambda\sigma}\nonumber\\+\frac{1}{4}C_{IJK}a^{J}a^{K}F^{0}_{4\mu\nu}F^{I}_{4\lambda\sigma}
\epsilon^{\mu\nu\lambda\sigma}+\frac{1}{2}C_{IJK}a^{K}F^{I}_{4\mu\nu}F^{J}_{4\lambda\sigma}
\epsilon^{\mu\nu\lambda\sigma}),\nonumber\\
\mu,\nu,\lambda,\sigma=t,r,\theta,\phi,~~~I=1,2,3.
\end{eqnarray}

Next, define
\begin{equation}
F_{0}\equiv e^{0}\frac{\partial f_{0}}{\partial
e^{0}}+e^{I}\frac{\partial f_{0}}{\partial e^{I}}-f_{0},
\end{equation}
where $F_{0}$ is a function of $v_{1}$, $v_{2}$, $w$, $a^{I}$ and
$x^{I}$. Finally, the entropy is given by
\begin{equation}
S_{BR}=2\pi F_{0}
\end{equation}
after extremizing $F_{0}$ with respect to various moduli and
substituting their values back into $F_{0}$.

We can obtain the explicit results of $\mathcal{L}^{'}_{0}$ and
$\mathcal{L}_{0CS}$ directly by putting the near horizon field
configuration~(\ref{3eq1}) into~(\ref{3eq2}),
\begin{eqnarray}
\mathcal{L}^{'}_{0}&=&(-\frac{2w}{v_{1}}+\frac{2w}{v_{2}}+\frac{(e^{0})^{2}w^{4}}{2v_{1}^{2}})\nonumber\\
& &+\frac{1}{2}(x^{I})^{-2}\frac{w^{2}}{v_{1}^{2}}
(e^{I}+a^{I}e^{0})^{2}-\frac{1}{8}(x^{I})^{-2}\frac{w^{2}}
{v_{2}^{2}}(p^{I})^{2},\nonumber\\
\mathcal{L}_{0CS}&=&2e^{-1}\sin\theta
C_{IJK}(e^{0}p^{I}a^{J}a^{K}+e^{I}p^{J}a^{K}).
\end{eqnarray}
One subtle is that the definition of electric charges will receive
modifications in the presence of Chern-Simons terms. In the usual
analysis of entropy function, the electric charges are defined by
$q_{I}\equiv\partial f/\partial e^{I}$. However, when Chern-Simons
terms are taken into account, the corresponding expression gives the
so-called ``Page charge'' introduced in~\cite{page}, whose
definition is given by
\begin{equation}
Q_{Page}\sim\int\ast F+A\wedge F.
\end{equation}

The subsequent calculations are straightforward and the expression
for $F_{0}$ is
\begin{eqnarray}
F_{0}&=&\frac{v_{2}}{2}-\frac{v_{1}}{2}+\frac{p^{2}L^{4}v_{1}}{32v_{2}w^{3}}
+\frac{1}{8}(x^{I})^{-2}\frac{w^{2}}{v_{1}^{2}}
(e^{I}+a^{I}e^{0})^{2}\nonumber\\&
&+\frac{1}{32}(x^{I})^{-2}\frac{w^{2}} {v_{2}^{2}}(p^{I})^{2},
\end{eqnarray}
where we have replaced $e^{0}$ by the ``true'' electric charge
$q_{0}\equiv\partial f^{'}_{0}/\partial
e^{0}=\frac{1}{4}v_{1}^{-1}v_{2}w^{3}e^{0}$. We can obtain the
correct values of the various moduli fields by solving the following
equations
\begin{equation}
\frac{\partial F_{0}}{\partial v_{1}}=\frac{\partial F_{0}}{\partial
v_{2}}=\frac{\partial F_{0}}{\partial w}=0,~~~\frac{\partial
F_{0}}{\partial x^{I}}=\frac{\partial F_{0}}{\partial a^{I}}=0.
\end{equation}
The solutions to the above equations are given as follows
\begin{equation}
\label{3eq9}
v_{1}=v_{2}=\frac{Lp^{2}}{4},~~~w=L,~~~x^{I}=\frac{p^{I}}{p},~~~a^{I}=-\frac{e^{I}}{e^{0}},
\end{equation}
note that the $x^{I}$s are not independent, subject to the
constraint $x^{1}x^{2}x^{3}=1$. Thus we have obtained the correct
near horizon geometry and attractor values of the scalar fields.
Furthermore, we can obtain the entropy by putting all the values
back into $F$,
\begin{equation}
S_{BR}=2\pi F_{0}=\frac{\pi Lp^{2}}{4},
\end{equation}
which reproduces the Bekenstein-Hawking entropy.
\subsection{Off-Shell Analysis}
\label{off}
We will calculate the entropy function using the off-shell
formalism. One advantage of the off-shell formalism is that the
supersymmetric completion of an $R^{2}$ term can be realized more
conveniently. For simplicity, we just list the basic ingredients of
the relevant supermultiplets briefly. Details for the
five-dimensional off-shell supergravity can be found in~\cite{hot}
and references therein. Similar work has been done in~\cite{fla},
where both the entropy function formalism and the so-called
``c-extremization''~\cite{pkfl} are discussed.

The irreducible Weyl multiplet, which consists of 32 bosonic plus 32
fermonic component fields, contains the following fields
\begin{equation}
e_{\mu}^{a},~~~\psi^{i}_{\mu},~~~V^{ij}_{\mu},~~~b_{\mu},~~~v^{ab},~~~
\chi^{i},~~~D,
\end{equation}
where $e_{\mu}^{a}$ are the vielbein, $V^{ij}_{\mu}$ and $b_{\mu}$
denote gauge fields associated with the $SU(2)$ generator and
dilatation generator respectively. $\psi^{i}_{\mu}$ and $\chi^{i}$
are $SU(2)$-Majorana spinors. Note that $v^{ab}$, $\chi^{i}$ and $D$
are auxiliary fields, where $v^{ab}$ is antisymmetric in $a$ and $b$
and $D$ is a scalar. The vector multiplet consists of gauge fields
$A^{I}_{\mu}$, scalar fields $M^{I}$, $SU(2)$-Majorana gaugini
$\Omega^{I}$ and $SU(2)$-triplet auxiliary fields $Y^{IJ}$, which
can be gauged away.

After gauge fixing, the bosonic terms in the two-derivative
Lagrangian of $\mathcal{N}=2$ supergravity with the Weyl multiplet
and $n_{v}$ vector multiplets can be expressed as
\begin{eqnarray}
\mathcal{L}_{0}&=&-\frac{1}{2}D+\frac{3}{4}R+v^{2}+\mathcal{N}(\frac{1}{2}D
+\frac{1}{4}R+3v^{2})+2\mathcal{N}_{I}v^{ab}F^{I}_{ab}\nonumber\\
               & &+\mathcal{N}_{IJ}(\frac{1}{4}F^{I}_{ab}F^{Jab}+\frac{1}{2}
               \partial_{a}M^{I}\partial^{a}M^{J})+\frac{1}{24}e^{-1}C_{IJK}
               A^{I}_{a}F^{J}_{bc}F^{K}_{de}\epsilon^{abcde},
\end{eqnarray}
where the functions characterizing the scalar manifold are defined
as
\begin{equation}
\mathcal{N}=\frac{1}{6}C_{IJK}M^{I}M^{J}M^{K},~~~
\mathcal{N}_{I}=\partial_{I}\mathcal{N}=\frac{1}{2}C_{IJK}M^{J}M^{K},~~~
\mathcal{N}_{IJ}=C_{IJK}M^{K},
\end{equation}
with $I,J,K=1,\cdots,n_{v}$. Note that the equation of motion for
the auxiliary field $D$ fixes $\mathcal{N}=1$, that is, the scalars
parametrize the ``very special geometry''. The auxiliary fields
$v^{ab}$ and $D$ can be eliminated via their equations of motion and
the resulting Lagrangian is the familiar one arising from the
compactification of eleven-dimensional supergravity on a Calabi-Yau
manifold with intersection numbers $C_{IJK}$.

The near horizon field configuration can be taken as follows
\begin{eqnarray}
\label{3eq14} &
&ds^{2}=w^{-1}[v_{1}(-r^{2}dt^{2}+dr^{2}/r^{2})+v_{2}(d\theta^{2}+\sin^{2}\theta
d\phi^{2})]+w^{2}(d\psi+e^{0}rdt)^{2},\nonumber\\
&
&F^{I}_{5rt}=e^{I}+e^{0}a^{I},~~~F^{I}_{4rt}=e^{I},~~~F^{0}_{4rt}=e^{0}=\frac{p}{2L},~~~
F^{I}_{5\theta\phi}=F^{I}_{4\theta\phi}=\frac{1}{2}p^{I}\sin\theta,\nonumber\\
& &M^{I}=mp^{I},~~~v_{\theta\phi}=V\sin\theta.
\end{eqnarray}
The auxiliary fields can be eliminated by solving the equations
\begin{equation}
\frac{\partial\mathcal{L}_{0}}{\partial
D}=0,~~~\frac{\partial\mathcal{L}_{0}}{\partial
V}=0,~~~\frac{\partial\mathcal{L}_{0}}{\partial m}=0,
\end{equation}
which gives
\begin{equation}
\label{3eq16}
D=12p^{-2},~~~m=p^{-1},~~~V=-\frac{3}{8}p.
\end{equation}
After substituting~(\ref{3eq16}) back in to $\mathcal{L}_{0}$, which
gives
\begin{eqnarray}
\mathcal{L}_{0}&=&(-\frac{2w}{v_{1}}+\frac{2w}{v_{2}}+\frac{(e^{0})^{2}w^{4}}{2v_{1}^{2}})-
\frac{3w^{2}}{8v_{2}^{2}}p^{2}-\frac{1}{2}c_{IJK}(e^{I}+e^{0}a^{I})(e^{J}+e^{0}a^{J})
\frac{p^{K}}{p}\frac{w^{2}}{v_{1}^{2}}\nonumber\\&
&-2e^{-1}\sin\theta C_{IJK}(e^{0}p^{I}a^{J}a^{K}+e^{I}p^{J}a^{K}).
\end{eqnarray} Then the subsequent analysis is similar to
the on-shell case discussed in the previous subsection and the same
result will be obtained, which would not be repeated here.
\section{Higher Order Corrections}
\label{HOC}
In this section we would like to discuss higher order corrections to
black ring entropy, which can be obtained in a similar way by
incorporating the higher order corrections into the effective
action. We will use two different actions, one of which is the
five-dimensional Gauss-Bonnet term coming from the compactification
of M-theory on a Calabi-Yau three fold $CY_{3}$~\cite{cor}, while
the other is a supersymmetric completion of $R^{2}$ terms in
five-dimensional supergravity proposed recently~\cite{hot}. We also
compare our results with the one obtained from microscopic
considerations.
\subsection{5D Gauss-Bonnet Corrections}
\label{5dgb}
The higher order corrections to the low energy effective action for
the compactification of M-theory on a Calabi-Yau threefold
$CY_{3}$(here is $T^{6}$) down to five dimensions takes the
following form
\begin{equation}
I_{GB}=\frac{1}{2^{9}3\pi^{2}}\int d^{5}x\sqrt{-g}c\cdot X
(R_{\alpha\beta\mu\nu}R^{\alpha\beta\mu\nu}-4R_{\alpha\beta}R^{\alpha\beta}+R^{2}),
\end{equation}
where $c\cdot X=c_{2I}X^{I}$ with $c_{2I}$ denoting the components
of the second class of $CY_{3}$. In principle, we should do
dimensional reduction on $\mathcal{L}_{GB}$ down to four-dimensional
spacetime. However, as pointed in previous section, such a term can
be evaluated in the original five-dimensional background without the
explicit resulting four-dimensional action, for this term is gauge
and coordinate invariant.

Define
\begin{equation}
f_{1}\equiv\frac{1}{16\pi}\int d\theta
d\phi\sqrt{-g}\mathcal{L}_{GB}
\end{equation}
and substitute the five-dimensional near horizon metric
in~(\ref{3eq1}) into the above expression, we can arrive at the
following result
\begin{equation}
f_{1}=\frac{1}{3\cdot2^{6}}\frac{c_{2I}p^{I}}{p}(-8w+\frac{2(e^{0})^{2}w^{4}}{v_{1}})
\end{equation}
where the scalar fields $X^{I}$ have been taken their attractor
values $X^{I}=p^{I}/p$. Then we can redefine $F\equiv F_{0}-f_{1}$
in a straightforward way and obtain the corrections to the entropy.
However, it seems that such equations can not give explicit
solutions for the moduli fields.

As pointed out in~\cite{ss2}, since the entropy is related to the
value of $F$ at its extremum, a first order error in the
determination of the near horizon background will give a second
order error in the value of the entropy. Thus if we want to obtain
the first order correction to the entropy, we can find the near
horizon background just by extremizing $F_{0}$ and then evaluate $F$
in this background. Finally, the entropy is given by
\begin{equation}
S_{BH}=2\pi F,
\end{equation}
after substituting the values of various moduli fields obtained by
extremizing $F_{0}$. Thus we can obtain the first order corrections
to black ring entropy
\begin{equation}
\label{4eq5}
\Delta S_{BR}=-2\pi f_{1}=\frac{\pi}{16}c\cdot
p\frac{L}{p}.
\end{equation}

\subsection{Supersymmetric $R^{2}$ Corrections}
\label{src}
One shortcoming of the five-dimensional Gauss-Bonnet Lagrangian is
that it is not supersymmetric, so several terms relevant to the
entropy might be omitted. Fortunately, a supersymmetric completion
of $R^{2}$ terms in five-dimensional supergravity has been realized
recently in~\cite{hot} and several relevant applications have been
listed in~\cite{fla}. Thus one can revisit the higher order
corrections by making use of the supersymmetric $R^{2}$ action.

A particular higher order term has been taken into account
in~\cite{hot} and the supersymmetric completion has been realized
there. The particular term is the so-called mixed
gauge-gravitational Chern-Simons term
\begin{equation}
e\mathcal{L}_{CS}=-\frac{c_{2I}}{6\cdot2}A^{I}\wedge\textrm{Tr}(R\wedge
R)=\frac{c_{2I}}{3\cdot16}\epsilon_{abcde}A^{Ia}R^{bcfg}{R^{de}}_{fg},
\end{equation}
with $c_{2I}$ being the expansion coefficients of the second Chern
class, which comes from the anomaly arguments. The four derivative
supersymmetric completion of the Chern-Simons term has been given
in~\cite{hot} and the relevant terms for the calculation
are\footnote{The difference in the overall coefficients is due to
the different conventions of the five-dimensional Newton constant.}
\begin{eqnarray}
\label{2eq5}
\mathcal{L}_{1}&=&\frac{c_{2I}}{3}(\frac{1}{16}e^{-1}\epsilon_{abcde}
A^{Ia}C^{bcfg}{C^{de}}_{fg}+\frac{1}{8}M^{I}C^{abcd}C_{abcd}+\frac{1}{12}M^{I}D^{2}
+\frac{1}{6}F^{Iab}v_{ab}D\nonumber\\&
&-\frac{1}{3}M^{I}C_{abcd}v^{ab}v^{cd}-\frac{1}{2}F^{Iab}C_{abcd}v^{cd}
+\frac{8}{3}M^{I}v_{ab}{\hat{\mathcal{D}}}^{b}{\hat{\mathcal{D}}}_{c}v^{ac}\nonumber\\
& &+\frac{4}{3}M^{I}{\hat{\mathcal{D}}}^{a}v^{bc}
{\hat{\mathcal{D}}}_{a}v_{bc}+\frac{4}{3}M^{I}{\hat{\mathcal{D}}}^{a}v^{bc}
{\hat{\mathcal{D}}}_{b}v_{ca}-\frac{2}{3}e^{-1}M^{I}\epsilon_{abcde}v^{ab}v^{cd}
{\hat{\mathcal{D}}}_{f}v^{ef}\nonumber\\
& &+\frac{2}{3}e^{-1}F^{Iab}\epsilon_{abcde}v^{cd}
{\hat{\mathcal{D}}}_{f}v^{ef}+e^{-1}F^{Iab}\epsilon_{abcde}{v^{c}}_{f}
{\hat{\mathcal{D}}}^{d}v^{ef}\nonumber\\
& &-\frac{4}{3}F^{Iab}v_{ac}v^{cd}v_{db}-\frac{1}{3}
F^{Iab}v_{ab}v^{2}+4M^{I}v_{ab}v^{bc}v_{cd}v^{da}-M^{I}(v_{ab}v^{ab})^{2}),
\end{eqnarray}
with $C_{abcd}$ the Weyl tensor defined as
\begin{equation}
C_{abcd}=R_{abcd}-\frac{2}{3}(g_{a[c}R_{d]b}-g_{b[c}R_{d]a})+\frac{1}{6}Rg_{a[c}g_{d]b}.
\end{equation}
The double covariant derivative of $v_{ab}$ is given by
\begin{equation}
v_{ab}{\hat{\mathcal{D}}}^{b}{\hat{\mathcal{D}}}_{c}v^{ac}=v_{ab}\mathcal{D}^{b}
\mathcal{D}_{c}v^{ac}+\frac{2}{3}v^{ac}v_{cb}{R_{a}}^{b}+\frac{1}{12}v_{ab}v^{ab}R,
\end{equation}
where the superconformal derivative is related to the usual
derivative as ${\hat{\mathcal{D}}}_{\mu}=\mathcal{D}_{\mu}-b_{\mu}$.

For our background which satisfies $\hat{\mathcal{D}}_{a}v_{bc}=0$,
following~\cite{ss2}, the first order corrections to the black ring
entropy can be obtained by substituting~(\ref{3eq9})
and~(\ref{3eq16}) into
\begin{equation}
\Delta^{'}S_{BR}=2\pi F^{'}_{1}=-2\pi f^{'}_{1},
\end{equation}
where
\begin{equation}
f^{'}_{1}\equiv\frac{1}{16\pi}\int d\theta
d\phi\sqrt{-g}\mathcal{L}_{1}.
\end{equation}
The result is
\begin{equation}
\label{4eq12} \Delta^{'}S_{BR}=\frac{\pi}{8}\frac{c\cdot p}{p}L,
\end{equation}
which shows similar behavior as the Gauss-Bonnet corrections but
with different numerical coefficients.
\subsection{Comparison with the Microscopic Corrections}
\label{com}
Now we would like to compare our results with the microscopic
entropy obtained in previous papers~\cite{ghls, bk}. The microscopic
corrections reads
\begin{equation}
\Delta S_{BRmic}=\frac{\pi}{6}c_{2}\cdot
p\sqrt{\frac{\hat{q_{0}}}{C}},
\end{equation}
where\footnote{Note that $q_{Ithere}=\frac{1}{4}q_{Ihere}$,
$p^{I}_{there}=\frac{1}{2}p^{I}_{here}$}$\hat{q_{0}}$ denotes the
left-moving oscillator number and $C$ is related to the left-moving
central charge $c_{L}$ by $c_{L}=6C$. The expressions for the
quantities are given as follows
\begin{eqnarray}
\hat{q_{0}}&=&-J_{\psi}+\frac{1}{12}C^{AB}q_{A}q_{B}+\frac{c_{L}}{24},\nonumber\\
&=&-J_{\psi}+\frac{D}{4}+\frac{1}{16}(\frac{q_{1}q_{2}}{p^{3}}+\frac{q_{2}q_{3}}{p^{1}}
+\frac{q_{1}q_{3}}{p^{2}})-\frac{1}{256D}((p^{1}q_{1})^{2}+(p^{2}q_{2})^{2}+(p^{3}q_{3})^{2}),\nonumber\\
C^{AB}&=&(C_{AB})^{-1}=(C_{ABC}p^{A}p^{B}p^{C})^{-1},\nonumber\\
c_{L}&=&6C=6C_{ABC}p^{A}p^{B}p^{C}=\frac{3}{4}p^{1}p^{2}p^{3}.
\end{eqnarray}
One can see that
\begin{equation}
\hat{q_{0}}=\frac{1}{8}L^{2}p,~~~C=\frac{1}{8}p^{3},
\end{equation}
then
\begin{equation}
\Delta S_{BRmic}=\frac{\pi}{6}c\cdot p\frac{L}{p},
\end{equation}

Comparing~(\ref{4eq5}) and~(\ref{4eq12}) with the above microscopic
correction, we can find that all of them behave similarly up to a
numerical constant. Furthermore, the correction obtained from the
supersymmetric completion is more accurate than the Gauss-Bonnet
correction, which means that the former action contains more terms
that are necessary in considering the higher order corrections to
black ring solutions.
\section{Summary and Discussion}
\label{Summary}
In this paper, we have constructed the entropy function for
supersymmetric black rings in $U(1)^{3}$ theory from both on-shell
and off-shell perspectives, which can be seen as a concrete
realization of~\cite{gj}. We have found that after dimensional
reduction down to four-dimensional spacetime, the effective action
is gauge invariant and we can carry out the entropy function
analysis. We have reproduced the Bekenstein-Hawking entropy
precisely and have obtained the correct attractor values of the
scalar fields via the entropy function using both formalisms. Note
that when we set $Q_{1}=Q_{2}=Q_{3}$, $q_{1}=q_{2}=q_{3}$, the black
ring becomes the solution presented in~\cite{eemr1}, which implies
that the entropy function formalism can also be applied to those
cases.

The higher order corrections to the entropy have also been discussed
and two kinds of corrections have been considered. One is due to the
five-dimensional Gauss-Bonnet term which comes from $R^{4}$
corrections in eleven-dimensional M-theory compactified on $CY_{3}$
but is not supersymmetric itself, while the other arises from a
supersymmetric $R^{2}$ completion of the five-dimensional
supergravity. Unfortunately, we have found that although both of
them can give similar behavior as the microscopic result, the
numerical coefficients can not be reproduced successfully. However,
the result obtained via the supersymmetric $R^{2}$ completion is
more accurate than that obtained via the five-dimensional
Gauss-Bonnet term, which means that the former contains more
information about the corrections and there might still be some
relevant terms ignored.

\acknowledgments DWP would like to thank Hua Bai, Li-Ming Cao and
Hui Li for useful discussions and kind help. The work was supported
in part by a grant from Chinese Academy of Sciences, by NSFC under
grants No. 10325525 and No. 90403029.
\end{document}